\documentclass[useAMS,usenatNishiNishibib,usegraphicx]{mn2e}

\title[]{The spiral structure of the Galaxy revealed
 by CS sources and evidence for the 4:1 resonance}
 
 \author[L\'epine et al.]{J.R.D. L\'epine$^{1}$\thanks{
 E-mail: jacques@astro.iag.usp.br}, A. Roman-Lopes$^{2}$, Zulema Abraham$^{1}$, T.C. Junqueira$^{1}$,
 Yu. N. Mishurov$^{3,4}$\\
 $^{1}$Instituto de Astronomia, Geof\'isica e Ci\^encias Atmosf\'ericas, Universidade
 de S\~ao Paulo, Cidade Universit\'aria, S\~ao Paulo, SP, Brazil\\ 
 $^{2}$Departamento de Fisica, Universidade de La Serena, Cisternas 1200, La Serena, Chile\\
 $^{3}$ Space Research Department, Southern Federal University, 5 Zorge, Rostov-on-Don, 344090, Russia \\
$^{4}$ Special Astrophysical Observatory of Russian Academy of Sciences,
 Nizhnij Arkhyz, Karachaevo-Cherkesia, Russia}

\begin{document}



\maketitle

\label{firstpage}

\begin{abstract}

We present a map of the spiral structure of the Galaxy, as traced by
molecular CS emission associated with IRAS sources which are believed to be
compact HII regions. The CS line velocities are used to determine the   
kinematic distances of the sources, in order to investigate their 
distribution in the galactic plane. This allows us to use 870 objects 
to trace the arms, a number  larger than that of  previous studies based
on classical HII regions. The distance ambiguity of the kinematic distances,
when it exists, is solved by different procedures, including the latitude 
distribution and an analysis of the longitude-velocity
diagram. The well defined spiral arms are seen to be confined inside the corotation 
radius, as is often the case in spiral galaxies. We identify a square-shaped 
sub-structure in the CS map with that predicted by  stellar
orbits at the 4:1 resonance (4 epicycle oscillations in one turn
around the galactic center). The sub-structure is found at the expected radius, based on
the known pattern rotation speed and epicycle frequency curve.
An inner arm presents an end with strong inward curvature and intense star formation that
we tentatively associate with the region where this arm surrounds the extremity of the bar, as
seen in many barred galaxies. Finally, a new arm with concave curvature is found
in the Sagitta to Cepheus region of the sky.

\end{abstract}


\section{Introduction}
While the spiral structure of external galaxies is clearly traced by CO or
by HII regions, to obtain the equivalent of a "face-on" map of the 
spiral arms of our Galaxy is a more tricky task. Since the Sun is situated 
close to the middle plane of the disk, the arms are seen superimposed. In
 addition, due to the high visual extinction in the disk, it is difficult to obtain 
reliable distances to stars from their absolute and apparent magnitudes. To 
determine the distances of HII regions or molecular clouds  in
the galactic plane, the most used tool has been the kinematic method.
 In this method, the velocity of the object, obtained from observations of radio
 recombination lines or of molecular lines, is compared to the predicted velocity
as a function of distance, based on the rotation curve. The main problem that one
has to face is the distance ambiguity that affects the sources  situated 
within the solar circle, which represents a large  portion of the galactic disk.
In this Galactic zone the method gives two solutions, and some additional information 
is required to decide which is the correct one. Another source of uncertainty
is the distance scale itself, which depends on the particular choice of the 
kinematic model of the Galaxy.

One of the first maps of the spiral arms based on star formation regions covering
large part of the Galactic plane was published by Georgelin \& Georgelin in 1976;
in that work, the known HII regions were fitted by 4 main spiral arms. Since then,
new radio recombination line surveys have been  completed (eg. Caswell and Haynes,
1987, Downes et al. 1980,  Lockman, 1989, Wink
et al. , 1983) and many papers have been devoted to resolving the distance
ambiguity by searching for absorption lines at a velocity higher than that of the
source. The presence of such lines is a strong argument in favour of the distant
solution (Kuchar \& Bania, 1994, Kolpak et al., 2003, Busfield et al, 2006). 
New maps of the spiral structure, making use of a larger number of HII regions, 
have been produced (eg. Paladini et al., 2004, hereafter PDD, Russeil , 2003). 
A map based on molecular clouds was presented by Efremov (1998).
In spite of the progress made, it cannot be said that a perfect knowledge of 
the spiral structure has yet been reached. If one examines the maps of PDD and 
of Russeil, for instance, one can see that the spiral structure is poorly 
determined within about 2 kpc from the Sun, and also that  beyond the
Galactic center there are many HII regions assigned to inter-arm regions, 
instead of being on the arms that are drawn by the authors.
Furthermore, there are noticeable differences between the two maps; for
instance, in the anti-center direction, PDD shows one arm while Russeil proposes
two arms. Concerning the solar neighbourhood, it seems that the use
of Cepheids is a better approach to get a consistent picture; see eg. Majaess et al.
(2009)

One way to improve the description of the spiral arms
is to increase  the number of objects used to trace them. We were able to
make progress in this direction by using the CS survey of IRAS sources with
characteristics of ultra-compact HII regions of Bronfman et al. (1996,
hereafter BNM). It was shown by BNM that the CS sources are remarkable tracers of the
spiral arms, but no attempt to use them for the whole Galaxy had yet been made.

The use of CS sources also permits to improve the method of solving distance
ambiguities. We show that there is an empirical relation between absolute IRAS
flux of the sources and CS line intensities, which allows us to choose between the two
kinematic solutions with a relatively good probability. We also make extensive
use of the latitudes (objects with latitudes of more than 0.5$^o$ have high 
probability of belonging to the near solution) and of the longitude-velocity
diagram, which gives a solution to the distance ambiguity in many cases. 
Furthermore, we are conscious that real galaxies do not show pure logarithmic
spiral arms over very long extensions; the arms are often made of different spiral
segments with some angle between them. For this reason, we do not attempt to force
fitting of extended logarithmic spirals. Also, based on the comparison with
external galaxies, we take into account the fact that real arms have a finite
width. The knowledge of how the width of the arms (of the order of 1 kpc in the
radial direction) reflects in the width of the lines in the longitude-velocity 
diagram is an important concept. Sometimes, points that are distant in the 
longitude-velocity diagram are, in reality, close together in the galactic plane
and belong to a same arm. Finally, the values for the galactic parameters
R$_0$ and V$_0$, and the choice of the  rotation curve and of the LSR correction, 
play important roles in obtaining a consistent spiral pattern.

\section[]{Galactic parameters, and calibration of kinematic distances with Cepheids}

A number of authors report serious discrepancies with distances derived from
the kinematic method, which are sometimes found to be up to 2 kpc larger than those
measured by other means (see for instance the photometric distances of groups of OB stars 
measured by Blum et al., 2000, and  Figueredo et al., 2008, and the study of  Foster \& 
MacWiliiams, 2006). This problem can be minimized by adopting the short
scale of the Galaxy (R$_0$ =7.5 kpc) and a relatively large
rotation velocity V$_0$ of the LSR. If a large velocity scale is used (numbers are
given below), a given difference of velocity between the LSR and the source is 
reached in a shorter distance. Furthermore, taking into account the minimum in the rotation
curve that exists close to the Sun also contributes to better results, since 
again larger velocity differences can be reached on shorter distances. We next 
argue that these  hypotheses on the rotation curve are reasonable in light  of 
recent studies.

In the last decade or even earlier, many authors have adopted R$_0$ = 7.5 kpc
(Racine \& Harris, 1989, Reid, 1993, among others). This shorter Galactic scale,
compared to the IAU recommended 8.5 kpc value, is supported by VLBI observations
of H$_2$O masers associated with the Galactic center. More recently  infrared 
photometric  studies of bulge red clump stars resulted in R$_0$=7.52 $\pm$0.10 kpc
(Nishiyama et al. 2006), while astrometric and spectroscopic observations of the
star S2 orbiting the massive black hole in the Galactic center taken at the ESO VLT 
(Eisenhauer et al., 2005) gives 7.94 $\pm$ 0.42 kpc. Bica et al. (2006), revisiting
the distribution of globular clusters, determined R$_0$ =7.3 $\pm$ 0.2 kpc.

The shorter scale has often been taken jointly with a smaller V$_0$, about 190 kms$^{-1}$,
since the ratio V$_0$/R$_0$ as derived from Oort's constants (V$_0$/R$_0$ = A-B) had a widely
accepted value of 25 kms$^{-1}$kpc$^{-1}$. Note that V$_0$/R$_0$ is the same
as the angular rotation velocity of the LSR, hereafter denoted $\Omega$. An interesting
independent measurement of $\Omega$, which does not depend on  the Galactic rotation curve,
is that of Backer \& Sramek (1999), who measured the proper motion of Sgr A at the Galactic 
center, obtaining 6.18 $\pm$ 0.19 mas yr$^{-1}$, which is equivalent to 29.2 $\pm$ 0.9 
kms$^{-1}$kpc$^{-1}$. (in what follows, the units are always kms$^{-1}$kpc$^{-1}$).
The origin of the discrepancies in the determination of $\Omega$ seems to be the 
fact that the method of Oort's constants is only valid if the rotation curve is
quite smooth, which is not the case in the solar vicinity. Olling and Merrifield
(1998) verified that the Oort's constants A and B differ significantly from the
general $V_0 /R_0$ dependence, in the solar neighborhood.  Olling \& Dehnen (2004)
argued that the most reliable tracers of the "true" Oort's constants are the red
giants, and derived A-B = 33. Among recent results, Branham (2002) obtains
$\Omega$= 30.3, Fern\'andez et al. (2001),$\Omega$ =30, and Miyamoto and Zu (1998),
$\Omega$ = 31.5,  from the kinematics of OB stars, Meztger et al. (1998)
obtained 31 from Cepheid kinematics, and Mendez et al. (1999)  31.7 from the 
Southern Proper Motion Program. This discussion tells us that it is quite 
reasonable to adopt R$_0$ = 7.5 kpc and V$_0$ about 220 kms$^{-1}$. 

In a study of the epicycle frequency in the galactic disk, based on  the spatial 
velocities of open clusters, Lepine et al. (2008, hereafter LDM)  argued that it was necessary to 
take into account a narrow minimum in the rotation curve at about 8.5 kpc
(using R$_0$ = 7.5 kpc), to fit the epicycle frequencies (see that paper for
other references on the minimum; we now prefer 8.8 kpc).

The rotation curve  that we adopted  is quite flat (except for a local
minimum at about 8.8 kpc),  and is close to that derived by Fich, Blitz
\& Stark (1989). The curve is conveniently fitted by exponentials and 
a Gaussian  (units are kms$^{-1}$ and kpc): 
\vskip -0.3cm

$$V= 360~e^{-r/3.1 -0.09/r} + 270~e^{-r/80 -(3.4/r)^2}$$
$$-15~e^{-[(r-8.8)/0.8]^2} ~~~~    (1)$$

The first two components contain terms in (1/r) and (1/r)$^2$ inside the exponential function,
which produces a decrease of their contribution towards small radii. The last term
represents a Gaussian minimum. Since we do not use the curve for radii smaller than
2.5 kpc, the first term, associated with the bulge, is not important in the present 
context. The interpretation of a similar curve in terms of components of the Galaxy 
is given by Leroy and Lepine (2001). We avoid here any theoretical model or discussion
on the mass components of the Galaxy, since our interest is only to adopt the best
empirical curve. We note however that the minimum at 8.8 kpc is not inconsistent
with the presence of a Cassini-like ring empty of HI at 8.4 kpc 
(Amores et al., 2009). It is easy to show that if the gap ring is responsible
for the minimum in the rotation curve, the minimum should coincide with the
outer edge of the ring, not with its middle. 

\begin{figure}
\includegraphics[width=84mm]{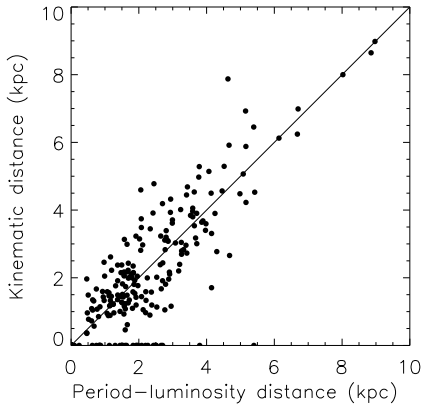}
\caption{kinematic versus photometric distances for Cepheid
stars} 
\label{figure1}
\end{figure}  

We present in Figure 1 a comparison of the kinematic and photometric 
distances of a sample of Cepheids. As the Cepheids are considered 
standard candles, the comparison is a test of the kinematic method.
The stars were taken from the catalog of Berdnikov et al.(2003), which
provides the distances based on the period-luminosity relation
and also lists the radial velocities of the stars. As a minor modification 
to the photometric distances, we re-computed them using the expression
for intrinsic colors as a function of period given by  Abrahamyan (2003)
and in addition, R$_v$=A$_V$/E(B-V)= 3.5 instead of the usual  3.1. This
improves the correlation and is justified by the fact that R$_v$ depends 
on the color of the stars (a correction must be made to the effective 
 wavelength of the B and V filters when they are convolved with the energy 
distribution of the stars (see eg. Azusienis \& Straizis, 1969). On the side 
of the kinematic calculation, we introduced a small expansion velocity
(1 kms$^{-1}$kpc$^{-1}$)in the galactic velocity field at large radii (R $>$ 8.5 kpc).
This also can be justified theoretically (the existence of the ring empty 
of HI revealed by Amores et al. is related to an outwards  flow of gas 
beyond corotation) but was introduced here empirically, as we found that 
this correction improves the correlation between the photometric and kinematic
 methods. The difference between the distances obtained by the two methods
is less than 0.7 kpc rms, and as can be seen in Figure 1; there is no
systematic discrepancy of one method with respect to the other, nor an increase
in the errors with distance. The errors on the kinematic distances
 are mostly explained by the non-circularity of the stellar orbits, 
while the errors on photometric distances are mainly due to the variety
of reddenning laws along the lines of sight to the stars. The interstellar 
extinction is often found to be anomalous, presenting values of R$_v$ larger
than 3.5 in the presence of molecular clouds along the line-of-sight. 
Figure 1 shows, however,  that there is no systematic difference between
the two methods, which could produce distorted or incorrect size of the spiral
structure.

We completed the CS sources with a short list (18) of methanol maser sources for which
precise measurements of parallax are available (Rygl et al., 2010, Sana et al., 2009,
and references therein). The methanol masers are associated with star forming
regions and are therefore  tracers of the spiral structure. 
The location of the masers  are shown in the figures in which 
we present maps of the CS sources on the galactic plane and are useful to confirm 
the position of the arms.

\subsection[]{Velocity perturbations with respect to the rotation curve}

The use of a well behaved or classical rotation curve like that of equation (1)
brings some difficulties. For a relatively large number of sources,
there are no kinematic solutions for the distances. For instance,
between galactic longitudes 28$^o$ and 33$^o$ the BNM catalog contains several 
tens of sources with LSR velocities of the order of 120 kms$^{-1}$. These
velocities are larger than the maximum velocity along the line-of-sight 
(or velocity at the subcentral point) derived from equation (1), for this
range of longitudes. A rotation curve with a much larger velocity at about 
6 kpc from the center would be required to obtain solutions in the usual way.
Such a rotation curve is unlikely on the basis of other observations 
including those at longitudes symmetrical with respect to the Galactic center.
The best explanation is that we are seeing
non-circular velocities. The way that we deal with this problem is the 
following: each time the computer program does not find a normal solution,
it adds (or subtract) a constant velocity (say, 20 kms$^{-1}$) to the theoretical 
curve of velocity versus distance, and looks again for a solution. In most 
anomalous cases a solution can then be found. Of course, a source with
anomalous velocity like in our example above cannot be far from the subcentral
point, and we could just attribute a distance equal to that of the subcentral
point. The solution that we adopted is preferable because it avoids crowding of
sources at the subcentral point, and smoothly distributes sources with different
velocities at different distances.

\section[]{Resolving the distance ambiguity}

Starting from the BNM catalog, and removing from it the sources 
with no CS detection, or with no IRAS identification, we get 852 sources.
We also remove 99 sources situated within $\pm 10\degr$ 
from the galactic center and from the anticenter, where kinematic distances  
are not valid; the remaining list contains 753 sources. Of these, 
for 236 sources there is only one kinematic solution, because they
are outside the solar circle (circle with radius R$_0$ centred on the 
Galactic center). For 500 sources two kinematic solutions were found
and for 16 sources no solution was found, even with the non-circular
velocity correction discussed in the last section. 
     
We next describe several criteria that we used to decide which of the 
two solutions is the correct one. Conflicts between different criteria 
were very rare. If more than one criterion points towards
the same (near or far) solution, then this solution has a high probability
to be correct.

A first very simple test is the galactic latitude. The CS sources are very
concentrated in the galactic plane, usually situated within 50 pc of the plane.
This means that a source with $\mid b\mid \geq 0.6^o$ has a large probability 
of being a nearby one. However, there are exceptions. In regions closer
than  5 kpc to the galactic center, the near and far solutions are
both far from the Sun and not very distant one from the other; in these cases,
this criterion looses its efficiency. Besides this, in some regions distant
from the galactic center, the warp of the Galaxy affects the average latitudes.
For instance, for longitudes 330$^o$-350$^o$, many sources have galactic latitudes
around -0.9$^o$; in these cases a distant (correct) solution is found at a distance
around 13 kpc from the Sun.

A second test makes use of the radioastronomical data on antenna temperatures
T and and line-widths $\Delta$V given in the CS sources catalog. It is reasonable
to suppose that there should be a correlation between the infrared and  radio
flux densities, but that possibly the angular size of the two components sources
are different and their ratio could depend on distance.
 We present in Figure 2 the integrated IRAS flux, defined 
in a simple way as the sum of $\lambda f(\lambda)$ for the 12 $\mu$m, 25 $\mu$m 
and 60 $\mu$m (the 100 $\mu$m band is excluded as it is too contaminated by
background radiation) versus T $\times \Delta$V of CS lines. In the figure, we plotted
the 236 sources with known distances (sources with only one
kinematic solution). The IRAS integrated flux is normalized to a distance of
1 kpc (in other words, it is the equivalent of an absolute flux).

Of course, we would expect to find intrinsically brighter sources among
the more distant ones, since weak sources at large distances would not be detected.
A surprise is that the distant sources are {\it much} brighter.
This has to be explained by the brightness distribution and space density
of the sources. As a comparison, let us remind that the list of the 25 stars 
nearest to the Sun and the list of the 25 brightest stars in the sky have
only 3 stars in common ($\alpha$ Centaurus, Sirius and Procyon). In Figure
2 we plotted with different symbols sources that are closer and farther
than 5 kpc, and an empirical separating line was drawn. Below the line the sources
have 95\% probability of being closer than 5 kpc. This is a distance which, for the 
sample with two solutions for the distance, is most often situated between the 
nearby and the distant solution. In practice, this criterion can be used in the following
way: we compute the distance-corrected flux for both the near and the far solution,
if both are below the line shown in Figure 2, then the near solution is very
probably the correct one.
  
\begin{figure}
\includegraphics[width=84mm]{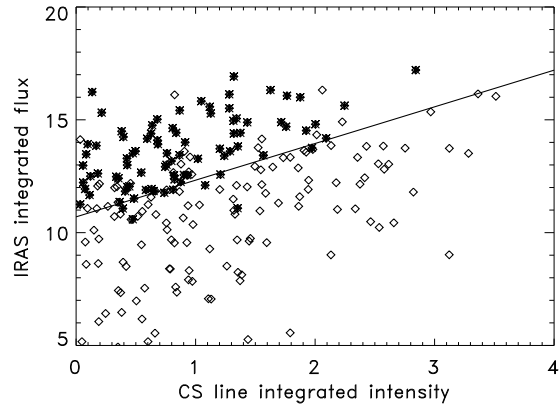}
\caption{Log of IRAS integrated, distance-corrected flux versus CS line intensity
(T$\times \Delta$V ) for sources with known distances. Sources situated at distances
larger than 5 kpc are filled symbols} 
\label{figure2}
\end{figure}  

Another criterion to choose the correct solution for CS sources is to 
look for the coincidences with HII regions for which the distance ambiguity
has already been solved. It is not rare that the CS sources have coordinates 
and velocities close to those of an HII regions, so that they probably 
belong to the same molecular cloud. For a number of HII regions
(see references in the Introduction) some criterion like the presence 
or absence of an HI absorption line at a high velocity was able to point
to the correct solution. 

Finally, we also made  use of the longitude-velocity diagram.
This diagram is convenient, since all the  sources
can be placed on it without problems of ambiguity.
Let us first remark that spiral arms in the first and fourth quadrants of 
the galactic plane, and situated at distances smaller than R$_0$ from the Sun,
present "tangential points", that is, a line-of-sight tangential to them 
can be drawn. These spiral arms have corresponding locii in the $l-v$ diagram
in the form of loops that reach a maximum value of longitude and then go back to
smaller values. For this reason all the sources that are
close to the lines of maximum velocities in the l-v diagram, that is, close to
 a line going from l = 90, v=0 to about  l=20, v=140, for positive longitudes, 
and to a line from l= -20, v=-140 to l=-90, v=0 (see Figure 3), are sources  
near a tangential direction to a spiral arm.  
For sources situated near the tangential directions, the near and far solutions
are close together, so that the problem of the distance ambiguity is 
not important. On the contrary, for sources that are about half way between
theses limiting lines and the center of the diagram (l=0, V=0), the two solutions
are distant.
 
\begin{figure}
\includegraphics[width=84mm]{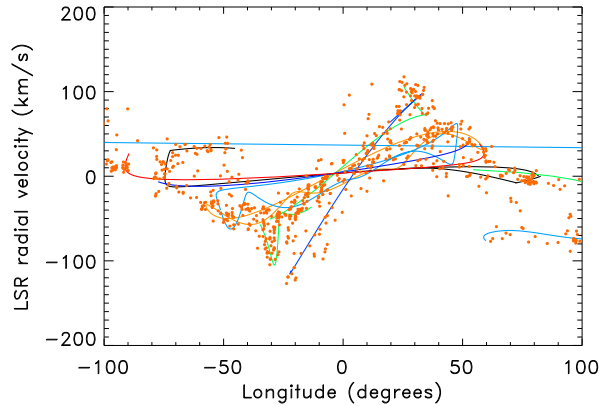}
\caption{The l-v diagram of the CS sources, and the locii  of the spiral arms
that we fitted to the spiral structure in Figure 7. We adopted the same colours
for the arms in the two figures, to turn the recognition easier .} 
\label{figure3}
\end{figure}  

\begin{figure}
\includegraphics[width=104mm]{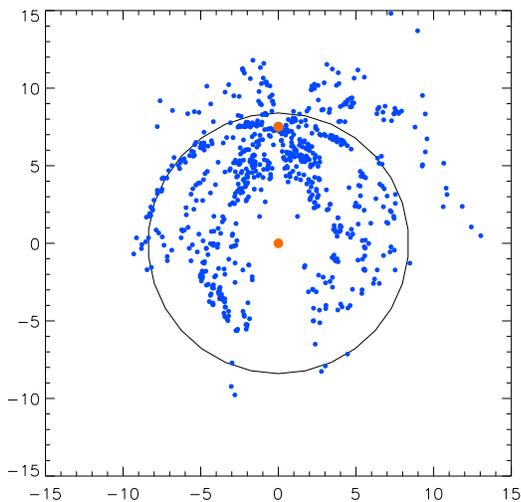}
\caption{The distribution of CS sources in the galactic plane.
The corotation circle and the positions of the Sun and of the Galactic center
are indicated.} 
\label{figure4}
\end{figure}

\section{Results and discussion}

Figure 4 shows the final result concerning the mapping of the spiral
structure with CS sources,  without any attempt to fit the arms.
Basically, our results confirm the main aspects of the spiral structure 
revealed by the studies of HII regions by Russeil (2004) and PDD.
For instance if we move horizontally across  the figure, 
to the right or to the left of the Galactic center, we find roughly 3
spiral arms on each side, like the previous works.  There are departures 
from the pure logaritmic spirals, with segments of arms that are almost
straight lines, for instance in directions around longitudes 320$^o$
and 30$^o$, which can also be seen in the Figure 12 of PDD. In the regions close
to the Sun the situation is not clear at all and, similarly to what
can be seen in Figure 3 (bottom) of Russeil, there are structures
that are oriented in directions which are not compatible 
with well behaved spirals. 

It is interesting to note that some of the spiral arms defined by the CS 
sources are  thinner than about 0.5 kpc in the radial direction over 
extensions of several kpc (see for instance the Carina arm, which coincides
with the corotation circle on the left side of the figure). In this sense 
he CS sources seem to be better arm tracers than the HII regions.  A possible 
reason for the better quality of the CS  map 
is that  the velocities of HII regions are obtained from radio recombination
lines, which are broad lines originating in ionized gas regions that are affected
by asymmetrical expansion and "champagne flow" effects, while the CS lines are
narrow. As we next discuss, the quality of the CS sources as tracers
allow us to distinguish substructures that were not  previously recognized.

\subsection{The role of the corotation radius}
At present the radius of corotation $R_c$ of the Galaxy is no more a question
of debate, since it has been determined by many independent
methods. We shall mention here two of them, which are particularly
precise: 1) the recovery of the birthplaces of  open clusters 
(Dias \& L\'epine, 2005), which allows us to see the change in position
of the arms as we compare different ranges cluster ages, the result
being $R_c$ = 1.06 $\pm$ 0.08 $R_0$, 2) the observation of a ring-shaped 
gap in the radial HI density distribution (Amores et al. 2009, hereafter ALM)
which was predicted by theory to occur at corotation , the 
average radius of the center of the ring being  8.3 kpc (for $R_0$ = 7.5 kpc).
Based on these results we adopt $R_c$ = 8.3 kpc. We show in Figure 4 the
position of the corotation circle. A natural question is: is there anything
special about it, in the spiral structure?
Elmegreem and Elmegreen (1995, hereafter EE95) commented in a work on the
morphology of spiral arms that in general, the arms are well defined (they
look like star-formation ridges) and are symmetrical inside corotation,
and are broader and more diffuse outside corotation. This corresponds 
well to what happens in our Galaxy, in which the well-defined arms
are almost all contained inside the corotation circle. Several 
"star formation ridges" seem to terminate at corotation. This is better seen
in Figure 8, which presents a zoom of the solar neighbourhood, where
two arms terminate at corotation.

\subsection{The fitted structure and the 4:1 resonance}

Before presenting the fits to the spiral structure, we must clarify what
we understand by spiral arms. The model that we adopt is based on ideas
of Kalnajs (1973). Stellar orbits of successive increasing radii in the disk are
organized in such a way that they get close together in some regions, which
consequently present an excess of stellar density.  The overdense zones form
elongated gravitational potential wells in the disk, which are the spiral
arms. The gas of the disk fall into the potential wells, reaching 
high densities that favours star-formation. Consequently,
the map of the star formation ridges coincides with the map of the regions
where the stellar orbits approach each other. The initial organization of
the stellar orbits is not a question to be debated here; it is 
probably due to the tidal effect of an external galaxy which passed close
to the Milky Way. A consequence of this interpretation of the physics 
of spiral arms is that the location of the arms is mainly determined 
by the stellar dynamics, not by the hydrodynamics of the gas.
The perturbed stellar orbits are not closed ones in the inertial frame
of reference of the Galaxy, but it is possible to find a rotating 
frame of reference in which the orbits are closed, so that stable
spiral arms are can be observed. The rotation velocity of that frame
is the pattern speed. A stellar orbit is closed if the epicycle frequency
(the frequency at which a star oscillates around the unperturbed circular
orbit) is a multiple of the angular frequency of the orbit around
the galactic center, in the pattern frame.

\begin{figure}
\includegraphics[width=84mm]{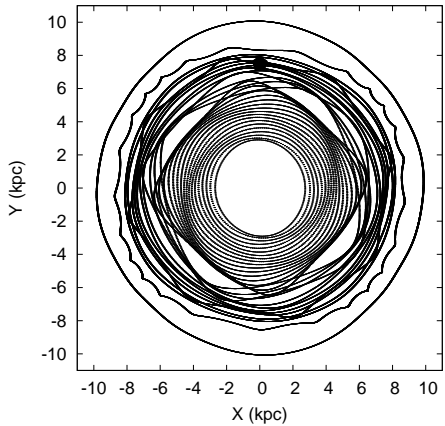}
\caption{A model of the spiral structure of the Galaxy based on the crowding
of stellar orbits to explain the arms.} 
\label{figure5}
\end{figure}

\begin{figure}
\includegraphics[width=84mm]{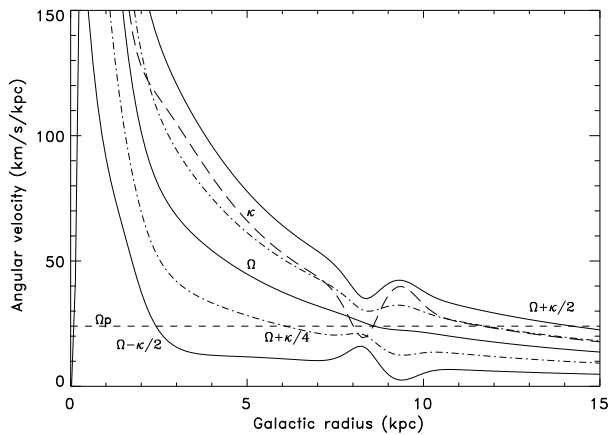}
\caption{The rotation curve of the Galaxy in angular velocity,
the epicycle frequency curve and the location of the main resonances,
for a pattern rotation velocity of 24 km/s/kpc. The $\Omega \pm \kappa/2$ curves 
are shown as coninuouos lines, and the $\Omega \pm \kappa/4$ curves as 
dotted-dashed lines. Corotation is at 8.4 kpc, ILB and OLB at 2.5 and 14 kpc
respectively, and the 4:1 resonance at 6.2 kpc.} 
\label{figure6}
\end{figure}

Amaral \& L\'epine (1997, hereafter AL) performed a study of the stellar 
orbits and spiral structure for a potential similar to that of our Galaxy.
We present in Figure 5 an updated version of that work; the method used
to trace the orbits is exactly the same of AL and does not need to be described here.  
Minor changes were introduced in some of the parameters like R$_0$ and the
rotation curve in order to be consistent with the present work. However,
we warn the reader that the model shown (at the present stage) is only
suitable for a qualitative understanding of the observed structure, but 
not for fitting the data.  

An interesting result is that the 4:1 resonance (4 epicycle oscillations
in 1 rotation) is prominent in the model and produces orbits that
look almost square-shaped. There is at this resonance a sudden 45$^o$ rotation 
between a square-shaped orbit and the next one at a slightly larger radius.   
    
Since in our Galaxy the corotation radius and the Inner Lindblad Resonance
(ILR, which we discuss later) are known, it is easy to infer were the 4:1
resonance is expected (see Figure 6): at about 6.2 kpc from the center.
This is the radius of the guiding center; if we consider a typical radial
amplitude of the epicycle oscillation of about 10\% of the radius of the
 orbit, the maximum distance from the galactic center
(the corner of the square) could reach about 7 kpc. Indeed, we claim 
that there is convincing evidence that this resonance is observed, as shown
in Figure 7. In that figure, we fitted square-shaped arms 
obtained by combining a circular orbit with radius 6.2 kpc plus an epicycle
perturbation with radial amplitude equal to 0.65 kpc (the radial "amplitude"
of the elliptic epicycle is the semi-minor axis $b$, with $a/b$ = 1.5,
see LDM). The phase of the epicycle
(or orientation of the square orbits) were adjusted to fit the data. It can be
seen that many CS sources fit the two sides of the rounded corner of the
square-shaped arm situated close to the Sun (in blue), as 
well as an entire side of the same square. The other square-shaped orbit
(in red), rotated 45$^o$ with respect to the first one, also presents
many CS sources along its sides.   Note that part of the corner close 
to the Sun, as well as the corner on the opposite side of the Galaxy,
are not observed because the sources with longitudes 10 degrees each side of
the Galactic center are not plotted (the kinematic method does not permit
to calculate the distances in those directions). Furthermore, the absence of sources
in distant regions opposite to the Galatic center, particularly at longitudes
between 330$^o$ and 350$^o$, is possibly due to observational limitations like the
limit of sensitivity of the IRAS survey based on which the CS sources were searched,
and or the warp of the Galaxy which favours negative longitudes (the survey was 
limited to $\pm$ 2$^o$ in these directions). 

\begin{figure}
\includegraphics[width=114mm]{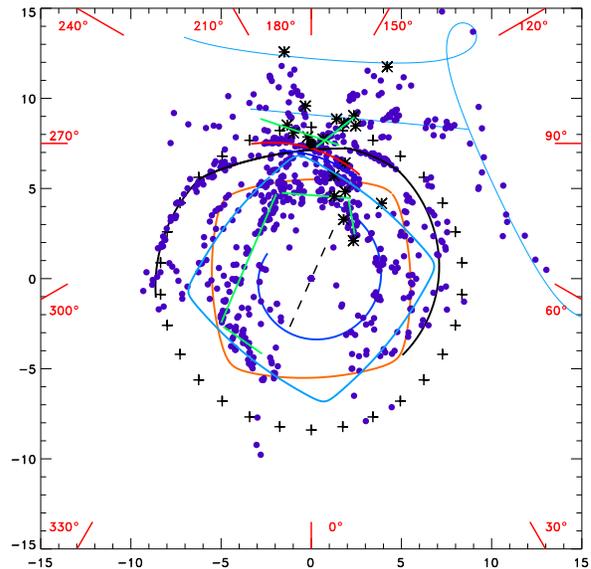}
\caption{The distribution of CS sources in the galactic plane, with fitted
square-shaped arms at the 4:1 resonance (in blue and in red) and segments
 of spirals and straight
segments at other locations. The corotation circle is indicated by crosses.
Masers sources with known distance from radio parallax  are indicated
as black stars. The outer scales of the box give distances to the Galactic
center in kpc; the inner scale in red gives the Galactic longitudes.} 
\label{figure7}
\end{figure}  

\begin{figure}
\includegraphics[width=104mm]{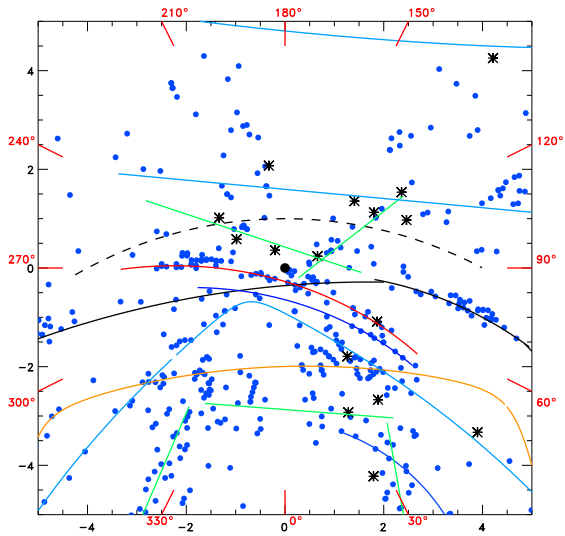}
\caption{The solar neighbourhood seen in an expanded scale. 
The colours of the fitted arms are the same of Figure 7. The CS sources and 
maser sources are shown.} 
\label{figure8}
\end{figure}

In order to overcome the problem of the absence of CS sources in a range of 
longitudes due to the kinematic distances, we also plotted the positions
of the Cepheids and of open clusters in the galactic plane, in Figures 9 and 10, 
similar to Figures 7 and 8. The sample of Cepheids and the method of calculation  of
their photometric distances were described in a previous section. The open clusters
were selected with an upper age limit of 15 Myrs (the catalogue is available at
$www.astro.iag.usp.br/\sim wilton$). The interstellar extinction limits the observation of
Cepheids and of open clusters to a small portion of the galactic plane, but these
objects are useful tracers in the solar neighbourhood. 
The Cepheids and clusters clearly delineate a part of the corner of the
square-shaped orbit at the 4:1 resonance which was lacking. The same rounded 90$^o$ break 
of that spiral arm was also observed by Majaess et al.(2009) using Cepheids.
Note that the amplitude of the radial perturbation is a parameter 
that strongly affects the shape of the corner, which can range from rounded
to a sharp peak, without affecting too much the rest of the square.

One may wonder why Cepheids, which are stars with ages up to 10$^8$ years, and 
CS sources associated with compact HII regions, with ages less than 10 million
years, could be sharing almost the same arm. Is it not expected that the stars
leave the spiral arms as soon as they form? This is true for "normal" logarithmic
spiral arms, since logarithmic spirals are not acceptable stellar orbits.
In the case of the 4:1 resonance  the square-shaped potential well, better
described as a potential channel or groove, is formed with the contribution
of many stellar orbits which are not square-shaped, as can be seen in Figure 5.
However, once the gravitational potential channel exists, the 
gas will flow along it, and the stars that form from that gas have initial
velocities directed along the direction of the groove, which means along the
square-shaped orbit. Since this coincides with an expected stellar orbit
in the galactic potential at that radius, the stars that are injected with the
correct velocity (magnitude and direction) will stay on that orbit. In a discussion
on initial velocities of open clusters LDM argued that the clusters tend to be
born with velocities in the direction of the arms. This is a mechanism that 
helps the arms to survive. In the same paper histograms of directions of velocities
of open clusters at their birth were presented for different ranges of galactic
radius, showing bimodal angular distributions. Such distributions can now be better 
understood in terms of the two directions in the resonant arm, before and after the corner. 

\begin{figure}
\includegraphics[width=104mm]{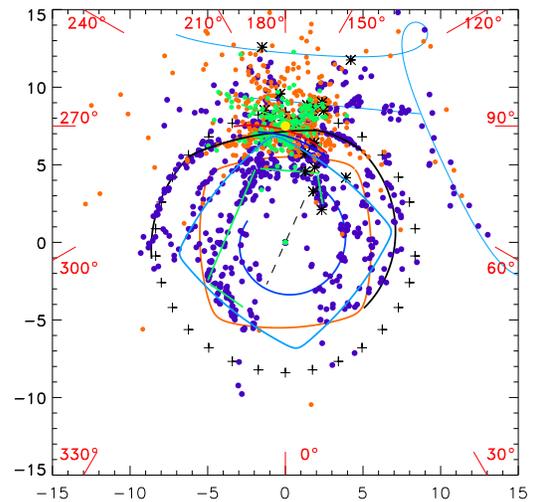}
\caption{The distribution of CS sources and maser sources in the galactic plane,
like Figure 7, plus the Cepheid stars in orange colour, and the young open clusters
in green. Here the Sun is represented by a yellow dot.} 
\label{figure9}
\end{figure}  

\begin{figure}
\includegraphics[width=104mm]{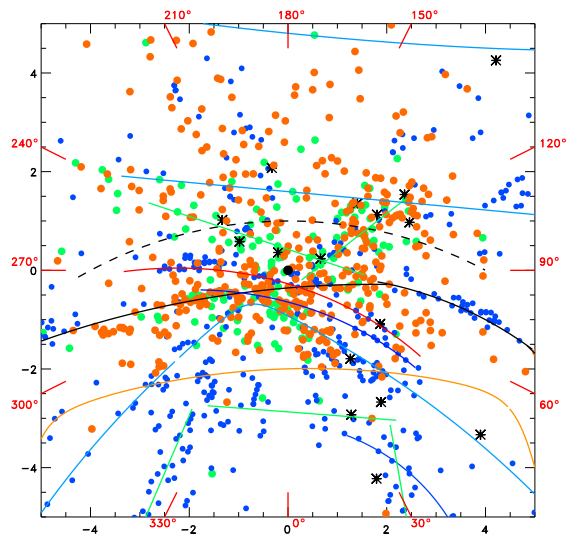}
\caption{The solar neighbourhood in an expanded scale of Figure 9, with the same objects
and same arms traced with the Cepheid stars in orange colour.} 
\label{figure10}
\end{figure}

\subsection{ The connection with the external structure}

Since we commented that several arms traced by the CS sources seem to
terminate at corotation, this may rise the question of wether the spiral
structure outside corotation is connected to that of the inner region.
There is clear evidence in other galaxies that the main arms are able
to cross corotation. For instance in M51 the corotation radius is
5 kpc or 120" (Sheepmaker et al., 2008); at this radius 
one can see a gap in the HI density in  one of the two main arms
(see the HI map of Walter et al. 2008), but the same arm seen in visible
light shows no discontinuity.

In our Galaxy too, we can see  connections  between the internal and external
spiral structures. For instance the Carina arm, which in our Figure 4  reaches
the corotation circle at about galactic longitude $\ell$  = 300$^o$, can be seen in 
Figure 2 of Levine and Blitz (2006, hearafter LB) to have a natural continuation
in an HI arm which can be followed up to $\ell$ = 320$^o$ (note that 
the circle represented by LB in their figure is the solar circle,
not corotation). An arm mapped by V\'azquez et al.(2008) in the third quadrant
of the Galaxy, seems to be a continuation of the arm that passes less than 1 kpc 
from the Sun in the direction of the anticenter (fitted by a green straight line in
Figure 10).

A relatively large ensemble of young objects containing methanol masers, Cepheids 
and open clusters, but only a few CS sources, is present in the direction 
of Perseus, l= 130$^o$, just outside corotation. It seems to be connected
with the structures in the inner side of corotation, as we indicate with a green 
straight line in Figure 10. For the moment it is difficult to say if this structure
is part of the outer arm that we show in blue, or if it is a "loop" in stellar
orbits (we comment these structures in a next section).   

In our Galaxy, only for some of the arms there seems to be  insufficient 
gas density to form stars near the "Cassini" gap in the gas distribution observed 
at corotation; this does not necessarily mean that the  grooves in the gravitational 
potential produced by stellar orbits are absent. Note that the investigations 
of stellar orbits usually skip the radii close to corotation because the integration
time to obtain an orbit around the Galactic center becomes very long.

\subsection{The Sagitta-Cygnus outer arm with concave curvature}

An interesting feature appears in Figures 7 and 9, which is an arm at about 10 kpc
from the Galactic center, in the range of longitudes about 80$^o$ to 140$^o$. This
arm is traced not only by CS sources but also by the Cepheids, which is a confirmation
of the correctness of the distance scale. This feature was neither recognized
as an arm by PDD nor by Russeil (2003) but it is clearly seen in the HI maps of LB 
(taking into account the difference in scale). Part of it appears in the map of 
molecular clouds by Efremov (1998), and it is represented by Vall\'ee (2008) as a very long
logarithmic spiral extending from Norma to Cygnus.

 The particularity of this arm, also seen in the map of LB, is that it presents
 a reverse curvature. Instead of attributing its shape to an interaction with an 
external galaxy, we prefer to explain it as a normal arm shape associated 
with a resonance. Like inside corotation, as previously discussed, we expect 
to have again outside corotation, but still within the OLR,
a collection of closed orbits with a number of "corners". If the radial 
amplitude of the epicycle perturbation is a large fraction of the radius of the
unperturbed orbit, the corners appear as narrow peaks pointing outwards
or even small loops in which the stars reverse their direction 
of motion, in the frame of reference of the pattern. In these cases,
 in the regions between two corners the curvature of the orbit is concave.
 In regions relatively distant to the center, the axisymmetric potential
 of the Galaxy becomes shallow, and  small energy perturbations 
 produce larger radial amplitudes. The curve that we plotted on the 
 external arm is only an illustration of the type of curve that
 can be obtained, not a final fit to the data. We also plotted a straight segment
 to roughly fit the sources in the upper part of the figure, just outside
 corotation. In next section we comment on the existence of such structures 
 in other galaxies.
 
 \subsection{Additional comments on polygonal arms}

If one looks carefully, in many cases, what seems at first glance to be a
logarithmic spiral arm is better approached by a sequence of straight segments
with an angle between  them (examples are M101, NGC 1232, and  M51).
Chernin et al. (2001) presents a list of a large number of galaxies with such
"polygonal" structures, which include some of the nearest and best known spirals.
Chernin (1999) explain the straight portions as local flatenning of the 
shock fronts that happens because a flat shock is a more stable configuration 
than a curved one. Our interpretation is different, since we consider that the
shape of the arms is determined by stellar orbits, not by the gas. Our model
predicts that for large perturbation amplitude, a loop appears in the connection between two 
straight segments, like the one that we draw in Figure 9. Such loops
are observed (see eg. the CO map of M51 by Hitschfeld et al. 2009, in the lower left 
of their Figure 5, and the image of M 101 available at the site
apod.nasa.gov/apod/ap030310.html, with two straight segments which seem to cross 
each other and make the connection through a big loop, at the top of the figure).
We do not exclude the existence of such loops in the 4:1 orbits of our Galaxy 
(the Gould Belt?), but this is left for future work.

In M 101 and NGC 1232, one can see many parallel segments of arms; this is  a 
justification for using two parallel straight segments, in the upper part of Figure 9, 
instead of attempting to fit a logarithmic spiral.

Concerning square structures in particular, Patsis et al. (1997) concluded
from a study of the morphology of spiral arms and interarm regions in SPH models
that the 4/1 resonance generates a clear signature in galaxies, namely a bifurcation
of the arms typical for the morphology of normal, late-type, grand design spirals.
Square structures are observed for instance in the inner parts of M 51
(see the CO map of the inner region by Aalto et al., 1999) and of 
NGC 5247 (Grosbol \& Dottori, 2008, see their Figure 4).

\subsection{The connection to the bar}

In the same work on the morphology of spiral arms that we already mentionned,
EE95 state that the bars of spiral galaxies can be separated in two
types, the large bars and the small bars. The small bars terminate
near the ILR of the spiral structure. This is clearly the case for our Galaxy.
The spiral arms can be followed down to about 3 kpc from the center,
as can be seen in Figure 7, which is close to the ILR of the
spiral structure, as derived from the rotation curve and  pattern
speed (see Figure 6).

Many authors refer to the "molecular ring" of our Galaxy, to describe the fact
that there is a larger density of molecular clouds in the range  3 to 5 kpc frm
the center.  This expression may give the impression that there is, indeed,
a ring, which  would be a closed arm-like structure. It is possible
to fit a closed ellipse to the CS sources near the ILR, but this is not 
the best fit; segments of  spiral arms seem to be a better approach.
We tentatively present the position of the bar with 3 kpc each side
of the center in Figure 7. Near the bar extremity nearest to the Sun, one can see
a spiral arm ending in a curvature that surrounds the bar, 
like in many barred galaxies (see pictures of NGC 1300, NGC 1073, etc.).
The precise position of this part of the arm is confirmed by the 
presence of two maser sources with direct measurement of distance 
by parallax. This suggests that the extremities of the bar are attached to the spiral
structure like it happens in most barred spirals, and that possibly,
the rotation velocity of the bar is the same of the spiral pattern 
(about 25 km/s/kpc). For the moment the rotation velocity of the bar
is uncertain, since precise measurements of  velocities of stars
really associated with the bar are difficult due to the heavy interstellar
extinction, and the velocity anomalies of stars situated near to the Sun are
only affected by the local spiral structure. We recognize that
this question is still a matter of debate (eg. Gerhard, 2010).

\section{Conclusions}
The CS sources of the catalog of BNM revealed to be very good 
tracers of the spiral arm structure of the Galaxy, covering a large 
part of the disk and providing a clearer picture than the HII regions.
A number of details that were not previously recognized could be
observed in the present study. The arms inside the corotation circle are thin and 
well defined, looking like star formation ridges, while outside corotation
they become broader and not so well defined. This is a general property of
spiral galaxies which was observed by EE95.

We were able to recognize the square-shaped stellar orbits of
the 4:1 resonance, very similar to those that are predicted from an 
analysis of the distribution of closed stellar orbits, using a model
based on  the potential of our Galaxy. One of the corners of 
these orbits is found to be at about 1 kpc from the Sun. This
discovery is rich in consequences that have not yet been  explored. 
The model shows that there are orbits of different sub-structures overlapping
in the solar neighbourhood. The stars from different orbits are expected
to have different space velocities and possibly different metallicities,
as they were born at different galactic radii.

The 4:1 resonance is a fundamental structure of the disk, and its identification,
added to the recent discovery of the ring-shaped gap in the gas distribution
at corotation, provides a robust self-consistent understanding of the spiral
structure.  The measurements of fundamental quantities like
the pattern rotation velocity and the epicycle frequency curve, recently
performed by our group, are in good agreement with the location
of the main resonances that we advocate.

In the inner regions of the galaxy, we observed a structure which seems to be 
the inner end of a spiral arm that makes a turn at the extremity of the bar,
as seen in many galaxies. Since in addition to the CS sources there are two
masers with distances directly determined by parallax in this arm extremity,
the precise distance can possibly restrict the models of the bar. 

Another interesting result is the observation of an arm with concave
curvature at about 10 kpc from the Sun in the longitude range 
80$^o$ to 140$^o$. We call it the Sagitta-Cepheus outer arm, since it is
not coincident with the arm called Cygnus arm in the literature, usually said
to be beyond the Perseus arm. This new arm is inside the OLR and its shape is 
not an unexpected one from the models of stellar orbits, which produces
concave orbits for large enough epicycle perturbation amplitude.
 
The work was supported in part by the S\~ao Paulo State
agency FAPESP and by the Conselho Nacional de 
Desenvolvimento Cientifico e Tecnol\'ogico. A. Roman-Lopes thanks support
from ALMA-CONICYT Fund, under the project number 31060004, "A new astronomer 
for the Astrophysics Group, Universidad de La Serena", and from the Physics
Department of the Universidad de La Serena.

\bsp

\label{lastpage}

\end{document}